\documentclass[sigconf]{acmart}[]

\acmConference[ICSE 2024]{46th International Conference on Software Engineering}{April 2024}{Lisbon, Portugal}

\usepackage{xcolor}
\usepackage{enumitem}

\usepackage{algorithmic}
\usepackage{graphicx}

\usepackage{textcomp}
\usepackage{booktabs}
\usepackage{multirow}
\usepackage{enumitem}
\usepackage{xcolor}
\usepackage{url}
\usepackage{balance}
\usepackage[multiple]{footmisc} 
\usepackage[english]{babel} 
\usepackage{lipsum} 
\usepackage{multirow}
\usepackage[lined,ruled,linesnumbered]{algorithm2e}
\usepackage{array}
\usepackage{arydshln}
\usepackage{bm}
\usepackage{listings}
\usepackage{tcolorbox}
\usepackage[font=small]{caption}
\usepackage[normalem]{ulem}
\usepackage{float}
\usepackage{siunitx} 

\usepackage[caption = false]{subfig}
\algsetup{linenosize=\small}
\SetKwInput{KwInput}{input}                
\SetKwInput{KwOutput}{output}
\SetAlFnt{\small}
\algsetup{linenosize=\tiny}   

\def \tool {\textsc{CureFuzz} }
\newcommand{\toolname}{{\sc CureFuzz}\xspace}

\newcommand\ans[1]{
\noindent 
\fcolorbox{green!40!black}{green!5}{\noindent 
 \parbox{0.98\columnwidth}{\noindent  #1}}\\
}

\usepackage{hyperref}
\hypersetup{hidelinks}

\usepackage{caption} 
\usepackage{xcolor}
\usepackage{enumitem}
\usepackage{amsmath}

\copyrightyear{2024} 
\acmYear{2024} 
\setcopyright{rightsretained} 
\acmConference[ICSE '24]{2024 IEEE/ACM 46th International Conference on Software Engineering}{April 14--20, 2024}{Lisbon, Portugal}
\acmBooktitle{2024 IEEE/ACM 46th International Conference on Software Engineering (ICSE '24), April 14--20, 2024, Lisbon, Portugal}\acmDOI{10.1145/3597503.3639149}
\acmISBN{979-8-4007-0217-4/24/04}
\begin{document}
\title{Curiosity-Driven Testing for Sequential Decision-Making Process}

\author{Junda He}
\affiliation{%
  \institution{Singapore Management University}
  \streetaddress{Address}
  \city{Singapore}
  \country{Singapore}
  \postcode{Postal Code}
}
\email{jundahe@smu.edu.sg}

\author{Zhou Yang}
\authornote{Corresponding author.}
\affiliation{%
  \institution{Singapore Management University}
  \streetaddress{Address}
  \city{Singapore}
  \country{Singapore}
  \postcode{Postal Code}
}
\email{zyang@smu.edu.sg}

\author{Jieke Shi}
\affiliation{%
  \institution{Singapore Management University}
  \streetaddress{Address}
  \city{Singapore}
  \country{Singapore}
  \postcode{Postal Code}
}
\email{jiekeshi@smu.edu.sg}

\author{Chengran Yang}
\affiliation{%
  \institution{Singapore Management University}
  \streetaddress{Address}
  \city{Singapore}
  \country{Singapore}
  \postcode{Postal Code}
}
\email{cryang@smu.edu.sg}

\author{Kisub Kim}
\affiliation{%
  \institution{Singapore Management University}
  \streetaddress{Address}
  \city{Singapore}
  \country{Singapore}
  \postcode{Postal Code}
}
\email{kisubkim@smu.edu.sg}

\author{Bowen Xu}
\affiliation{%
  \institution{North Carolina State University}
  \streetaddress{Address}
  \city{Raleigh}
  \country{United State}
  \postcode{Postal Code}
}
\email{bxu22@ncsu.edu}

\author{Xin Zhou}
\affiliation{%
  \institution{Singapore Management University}
  \streetaddress{Address}
  \city{Singapore}
  \country{Singapore}
  \postcode{Postal Code}
}
\email{xinzhou.2020@phdcs.smu.edu.sg}

\author{David Lo}
\affiliation{%
  \institution{Singapore Management University}
  \streetaddress{Address}
  \city{Singapore}
  \country{Singapore}
  \postcode{Postal Code}
}
\email{davidlo@smu.edu.sg}

\begin{abstract}
Sequential decision-making processes (SDPs) are fundamental for complex real-world challenges, such as autonomous driving, robotic control, and traffic management. 
While recent advances in Deep Learning (DL) have led to mature solutions for solving these complex problems, SDMs remain vulnerable to learning unsafe behaviors, posing significant risks in safety-critical applications.
However, developing a testing framework for SDMs that can identify a diverse set of crash-triggering scenarios remains an open challenge. 
To address this, we propose \toolname, a novel curiosity-driven black-box fuzz testing approach for SDMs. \toolname proposes a curiosity mechanism that allows a fuzzer to effectively explore novel and diverse scenarios, leading to improved detection of crash-triggering scenarios. 
Additionally, we introduce a multi-objective seed selection technique to balance the exploration of novel scenarios and the generation of crash-triggering scenarios, thereby optimizing the fuzzing process. 
We evaluate \toolname on various SDMs and experimental results demonstrate that \toolname outperforms the state-of-the-art method by a substantial margin in the total number of faults and distinct types of crash-triggering scenarios. We also demonstrate that the crash-triggering scenarios found by \toolname can repair SDMs, highlighting \toolname as a valuable tool for testing SDMs and optimizing their performance.
\end{abstract}


\maketitle

\section{Introduction}
\label{sec:intro}
Sequential decision-making processes (SDPs) involve a series of interrelated decisions, where each decision depends on the outcome of the previous one. SDPs play a critical role in addressing various complex real-world challenges such as autonomous driving~\cite{isele2018navigating}, robotic control~\cite{kober2013reinforcement}, and traffic control~\cite{wu2017flow}.  
Recent advances in Deep Learning (DL), e.g., Deep Neural Networks (DNN)~\cite{schmidhuber2015deep}, Deep Reinforcement Learning (DRL)~\cite{SilverHMGSDSAPL16}, and Imitation Learning (IL)~\cite{abs-2001-02328}, have led to mature solutions for handling these complex sequential decision-making problems. We refer to these solutions as sequential decision-makers (SDMs). 
In various applications, such as video game playing~\cite{SilverHMGSDSAPL16,HosuR16,fuchs2021super}, and aircraft collision avoidance systems~\cite{acasxu}, these SDMs demonstrate human-comparable or even superior capabilities.

Despite impressive effectiveness, SDMs are susceptible to learning unsafe behaviors during the training process~\cite{MDPFuzz}. As SDMs primarily aim to optimize overall performance, they might not adequately prioritize safety concerns. This learning flaw could potentially lead to catastrophic failures in real-world scenarios~\cite{MDPFuzz, zhong2022neural}.
The impact of such risks is particularly severe in safety-critical applications, where the actions of SDMs directly affect human lives, and any negligence could result in disastrous consequences. 
A heartbreaking report\footnote{
\href{https://www.engadget.com/self-driving-car-technology-crash-data-172606258.html\#:~:text=As\%20The\%20New\%20York\%20Times,2021\%20and\%20May\%2015th\%2C\%202022}{www.engadget.com/self-driving-car-technology-crash-data}} reveals that between July 2021 and May 2022, there were 392 crashes involving autonomous vehicles, leading to five serious injuries and six deaths. 
This highlights the necessity to thoroughly test SDMs to ensure safety in critical scenarios before their deployment.


An ideal testing framework for SDMs must be able to produce a diverse set of crash-triggering scenarios~\cite{scenoRITA}. A scenario, in this context, is defined as the set of environmental properties with which an SDM interacts. For example, in autonomous driving, a scenario would include the description of pedestrians, other vehicles, and traffic conditions. The generation of a diverse set of crash-triggering scenarios is crucial for several reasons. Firstly, identifying duplicate (i.e., the opposite of diverse) crash-triggering scenarios wastes computational resources that could have been used to uncover new bugs. Secondly, less diverse scenarios cover smaller input space, thus uncovering fewer corner case crashes.

However, enhancing the diversity of crash-triggering scenarios is indeed challenging, as it necessitates a method to measure the novelty of a scenario. In the context of sequential decision-making problems, the state space is often high-dimensional and continuous. This makes the computation of novelty difficult and leads to high computational costs, commonly known as the ``curse of dimensionality''~\cite{koppen2000curse}. Thus, researchers are motivated to propose novelty measures for scenarios in the context of testing SDMs. Nonetheless, existing novelty measures have their limitations. For instance, the density-based method~\cite{MDPFuzz} exhibits high computational complexity, especially in environments with high complexity. Moreover, the topological similarity measurement proposed by Li et al.~\cite{diffusion} can only measure the novelty of the termination states. While typically SDMs make hundreds of interactions with the environment in each run, and the rich information from the intermediate states is buried. Consequently, finding a more efficient and effective novelty measure that helps in generating diverse test cases for SDMs remains a to be an ongoing challenge.

Inspired by the concept of Random Network Distillation (RND) in reinforcement learning~\cite{rnd}, we propose a curiosity mechanism and a novel fuzz testing method to address the aforementioned challenge. 
Our curiosity mechanism calculates the extent of the fuzzer's curiosity in exploring specific scenarios and encourages the fuzzer to prioritize scenarios with higher curiosity. 
Specifically, when presented with a scenario, our curiosity mechanism will predict the subsequent states within its environment. The discrepancy between our prediction and the actual outcome (prediction error) reflects the level of the fuzzer's curiosity. A scenario with higher curiosity suggests that it is unfamiliar to the fuzzer, as the fuzzer cannot accurately predict it. 
An essential advantage of our curiosity mechanism is its computational efficiency. The computational complexity of our curiosity mechanism increases linearly while the environmental complexity increases. 
By leveraging this approach, we can both effectively and efficiently encourage the fuzzer to explore uncharted territories, thereby increasing the diversity of scenarios generated. The computationally inexpensive feature also allows us to successfully apply our method to complex environments, including self-driving systems.

With the novel curiosity mechanism, we propose \toolname (\textbf{Cur}iosity-driv\textbf{e}n \textbf{fuzz}er), a curiosity-driven fuzz testing approach for Sequential Decision Making Process. \toolname combines two distinct techniques: (1)~a curiosity mechanism that measures the novelty of encountered scenarios and encourages the fuzzer to detect novel and diverse scenarios; (2)~a multi-objective seed selection technique in fuzzing that estimates the energy of a seed based on its probability of triggering crashes and exploring novel scenarios. \toolname then selects the seed for mutation based on its estimated energy to guide the search for crash-triggering scenarios. The former of these two techniques provides a diversity of generated scenarios, and the latter ensures effectiveness in detecting crash-triggering scenarios.
By combining them, \toolname achieves a balance between effectiveness and diversity in fuzzing testing of the Sequential Decision Making Process, leading to a significant improvement over the state-of-the-art.


We evaluate \toolname by applying it to well-known SDMs that use various learning algorithms.
The algorithms include Deep Neural Networks (DNN)~\cite{montavon2018methods}, Deep Reinforcement Learning (DRL)~\cite{arulkumaran2017deep}, Multi-agent DRL (MARL)~\cite{egorov2016multi}, and Imitation Learning (IL)~\cite{hussein2017imitation}.
We also consider a range of sequential decision-making problems. (i.e., autonomous driving~\cite{carlachallenge}, aircraft collision avoidance~\cite{acasxu}, and video game playing~\cite{LoweWTHAM17,kuznetsov2020controlling}). 
The experimental results demonstrate that \toolname effectively and efficiently identifies a significant number of catastrophic failures across all considered SDMs.
Overall, \toolname outperforms the state-of-the-art methods and detects a more diverse set of crash-triggering scenarios. 
Furthermore, we have also shown that the crash-triggering scenarios identified by \tool can be utilized to repair the SDMs. By re-running \tool on the repaired SDMs, the number of detected faults decreases by 73\%, highlighting the practical utility of our approach in enhancing the effectiveness of SDMs.

\noindent The contributions of this paper are summarized as follows:
\begin{itemize}[leftmargin=0.5cm]
    \item We introduce \toolname, the first curiosity-driven black-box fuzz testing framework for DL-based sequential decision makers. \toolname aims to reveal a diverse set of crash-triggering scenarios, enhancing the safety and effectiveness of these decision-making systems.
    \item We propose a novel curiosity mechanism that leverages the prediction error of two neural networks to measure the novelty of scenarios for fuzz testing. 
    \item To evaluate the effectiveness of \toolname, we conducted experiments on 4 sequential decision-making tasks. The results demonstrate that \toolname successfully uncovers crash-triggering scenarios and outperforms our baseline method by a substantial margin.
\end{itemize}

\section{Preliminaries}
\label{sec:background}




\subsection{Markov Decision Process}
The Markov Decision Process (MDP) is a well-known mathematical framework for modeling complex sequential decision-making problems under various uncertainties~\cite{puterman1990markov}. 
In this paper, we focus on the SDMs solving MDPs. 
An agent (i.e., defined as SDM) and the environment are the two main components of an MDP. In a general paradigm, an agent actively engages with its environment through a sequence of actions. Upon executing an action, the environment responds by transitioning to a new state and provides the agent with feedback in the form of rewards. This reward signal serves as a measure of the quality of the agent's action. The agent's ultimate objective is to learn an optimal policy, which is a strategy that guides its decision-making process to maximize the accumulated reward over time.
For simplification, the interactions are assumed to be performed in discrete time steps. 
Given $t= 0,1,2,...$ denotes the discrete time step, we formally define the Markov Decision Process as a tuple of $(S, A, T, R)$, where:
\begin{itemize}[leftmargin=*]
    \item $S$ is the state space, a set of states that represents all the possible statuses of the environment. A state $s \in S$ refers to the current situation of the agent, and $s_t$ refers to the state at time $t$. 
    \item $A$ is the action space, a set of available actions that can be taken by the agent. Given a state $s$, the agent selects its action $a \in A$, accordingly.
    \item $T$ is the state transition probability function of the environment, such that
     $T(s' | s, a)$ describes the probability of transitioning to $s'$ from $s$ when the action $a$ is taken.
    \item $R$ is the reward function. $R(s, a, s')$ refers to the immediate reward received by the agent when it takes the action $a$ at state $s$ and reaches $s'$.
\end{itemize}
A sequential decision-making problem is an MDP if and only if it satisfies the Markov Property. In other words, the decision-making process in this environment depends solely on the current state of the environment and not on the sequence of past states. Mathematically, \textbf{Markov Property} is described as:
\begin{small}
    $$P(S_{t+1} = s_{t+1} | S_t = s_t, ... , S_0=s_0)
    = P(S_{t+1} = s_{t+1} | S_t = s_t)$$
\end{small}
$P$ denotes the probability of transitioning from $S_t = s_t$ at time step $t$ to $S_{t+1} = s_{t+1}$ at time step $t+1$. Intuitively, the Markov property ensures that the current state encapsulates all relevant information from history and that the future of the process is independent of the past when the present state is known.


\subsection{Sequential Decision Makers}
Sequential Decision Makers (SDMs) powered by deep learning have shown strong capabilities in solving MDPs. 
Here, we introduce the technology used in four state-of-the-art DL-based SDMs:  Deep Neural Networks (DNN)~\cite{montavon2018methods}, Deep Reinforcement Learning (DRL)~\cite{arulkumaran2017deep}, Multi-agent DRL (MARL)~\cite{egorov2016multi}, and Imitation Learning (IL)~\cite{hussein2017imitation}.

\subsubsection{Deep Neural Network }
Although all four techniques involve neural networks, we refer to DNN~\cite{montavon2018methods} when the training process is conducted in the supervised setting. In this case, the optimal actions to take for certain states are labeled. During the training session, the model learns to derive the optimal policy by minimizing the difference between its predicted actions and the manually labeled actions. One application is the aircraft collision avoidance system, ACAS Xu~\cite{acasxu}. 

\subsubsection{Deep Reinforcement Learning}
DRL~\cite{MnihKSRVBGRFOPB15} combines the notion of deep learning with traditional reinforcement learning techniques. Instead of learning with labeled data, a DRL agent learns optimal policy through trial-and-error with the rewards or penalties received from the environment. Algorithms like DQN~\cite{MnihKSRVBGRFOPB15}, PPO~\cite{schulman2017proximal}, and TQC~\cite{kuznetsov2020controlling} have been used in a wide range of tasks. For example, AlphaGo~\cite{SilverHMGSDSAPL16}, and OpenAI's DOTA 2 agents~\cite{ye2020towards}. 

\subsubsection{Multi-agent Reinforcement Learning}
MARL extends conventional reinforcement learning to scenarios that have multiple agents.
MARL is a type of machine learning in which multiple agents learn to interact with each other in a shared environment through a feedback mechanism. In MARL, agents learn to make decisions also based on the actions of other agents in the environment. A MARL game can be cooperative, competitive, or a mix of both. In a cooperative game, the agents work together towards a common goal, where the success of one agent is dependent on the success of the other agents.

\subsubsection{Imitation Learning}
IL is widely used in scenarios where reinforcement learning may be too slow or expert demonstrations are available, such as autonomous driving.
IL usually involves two agents: an expert agent and a student agent. It can be considered as another form of supervised learning, where the student agent aims to mimic the behaviors of the expert agent. 

\subsection{Fuzz Testing}

Fuzz testing is a widely used method in software testing that automates the generation of inputs to identify vulnerabilities, crashes, or any other unexpected behavior in software programs. 
This approach usually creates a wide range of variants from a set of initial inputs (often referred to as seed corpus). These variants are generated by applying certain mutation operations, such as bit flipping~\cite{cha2015program}. The mutated inputs are then used to test the software, with the intent of causing unexpected behavior or crashes. The advantage of mutation-based fuzzing lies in its capability to explore the program execution paths that might not be covered under standard testing procedures, thereby increasing the robustness and security of the software.


\section{Approach}
\label{sec:method}
\begin{figure}[!t]
  \centering
  \includegraphics[width=1.0\linewidth]{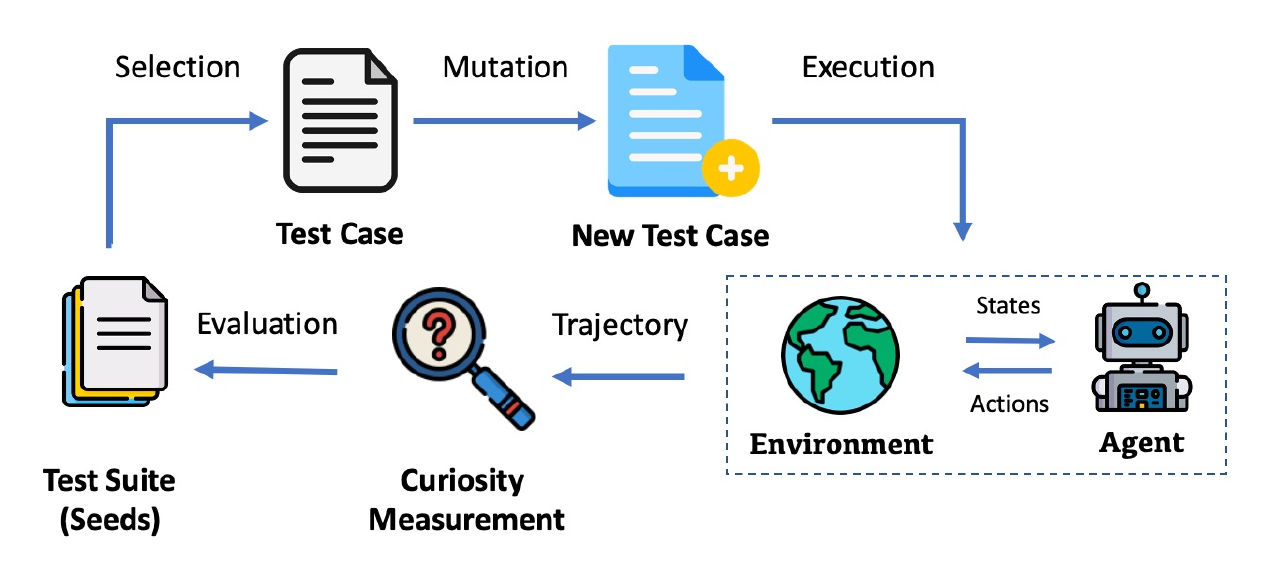}
  \caption{Illustration of the overall workflow of the proposed approach \toolname.}
  \label{fig:overview}
  \vspace{-10pt}
\end{figure}
\subsection{Assumption}
We focus on environments where the transition dynamics satisfy the Markov Property~\cite{puterman1990markov}, and our testing subjects are DL-based SDMs (agents) solving MDPs. \toolname performs fuzz testing in a black-box manner, which is realistic and practical. While white-box testing can be valuable for certain tests, the high complexity of deep learning models, potentially with millions of parameters, can make white-box testing overwhelming. Additionally, in the real world, SDMs' internals are often not available (e.g., users access the SDM through a third-party vendor's service). More specifically, \toolname does not require access to the SDM's internal logic, nor knowledge of the environment's transition dynamics or reward mechanism. Moreover, the SDM's policy remains fixed and will not be updated during the fuzzing process. Given a particular state, \toolname can obtain the corresponding action from the SDM by interacting with the environment.


\toolname is designed to uncover \textit{crash-triggering scenarios} that eventually lead to the crash of SDMs. It is crucial to note that the definition of a `crash' can vary across different environments. For instance, within the context of autonomous driving, a crash could be defined as an incident where an autonomous vehicle collides with pedestrians. In the context of robotics control, a crash can refer to the falling of a walking robot. We employ the term `crash' to represent its broader conceptual meaning. Furthermore, our methodology focuses on catastrophic failure rather than minor deviations from optimal performance. If an SDM initiates from a particular scenario, and the subsequent cumulative reward falls below a predefined threshold, yet no catastrophic failure is observed, we do not classify this as a failed case. 
\subsection{Approach Overview}

\toolname consists of the following stages: Setup and initialization, seed energy estimation, seed selection, seed mutation, and seed evaluation. 
Algorithm \ref{algo:workflow} presents the high-level workflow of \toolname and visually represented in Figure \ref{fig:overview}. As it core, \toolname maintains a corpus of seeds, each representing a unique scenario in the environment. We refer to the starting condition of a specific scenario as the initial state. We then observe the resultant actions of the SDM under the initial state.
The SDM's actions change the environment and lead to new states, thus we obtain the induced state sequence for each initial state. The interaction between the SDM and the environment is automatically terminated if the state sequence reaches its maximum length or a crash is detected.
\toolname then estimates the energy of a seed based on its intrinsic reward (to be described later in Section \ref{subsec:rnd}), the probability of triggering new scenarios (to be described later in Section \ref{subsec:architecture}), and the probability of triggering a crash (to be described later in Section \ref{subsec:architecture}). 
Seeds with higher energy, indicating either their novelty or higher crash probability, are selected preferentially for mutation. This mutation process produces new seeds, whose induced state sequences are then evaluated. This cycle repeats throughout the fuzz testing procedure.

\begin{algorithm}[!t]
    \DontPrintSemicolon
    \caption{The Curiosity Mechanism}
    \label{algo:rnd}
    \SetKwFunction{Curiosity}{Curiosity}
    \SetKwFunction{initCuriosity}{initCuriosity}
    \SetKwProg{Fn}{Function}{:}{}

    \Fn{\initCuriosity{}}{
      Initialize target network $\phi_{target}$\;
      Initialize predictor network $\phi_{pred}$ with same architecture as $\phi_{target}$\;
      Fix the weights of $\phi_{target}$ to random values\;

      return $\phi_{target}$, $\phi_{target}$

    }

    Note that  $\boldsymbol{S} = \{s_1, s_2, ..., s_t\}$\

    \Fn{\Curiosity{$\boldsymbol{S}$, $\phi_{target}$, $\phi_{target}$}}{
      $intrinsic\_reward = 0$

      \For{$i=1$ to $t$}{
        \tcc{Compute intrinsic reward
        }
           $r_{i} = \|\phi_{target}(s'_{i}) - \phi_{pred}(s'_{i})\|^2$\; 

           $intrinsic\_reward += r_{i}$\;
           \tcc{Update predictor network
           }
           Update $\phi_{pred}$ to minimize $\|\phi_{target}(s'_{i}) - \phi_{pred}(s'_{i})\|^2$\;      
           }
           $intrinsic\_reward = intrinsic\_reward \times \frac{1}{t}$

        \KwRet $intrinsic\_reward$, $\phi_{target}$, $\phi_{target}$\;
        }
    
    \end{algorithm}

\subsection{A Curiosity-driven Search Strategy}
\label{subsec:rnd}
Inspired by the success of exploration engineering~\cite{li2017deep}, we propose a curiosity mechanism that measures the novelty of a scenario. 
Algorithm~\ref{algo:rnd} shows the pseudocode for our curiosity mechanism. Our curiosity module consists of a pair of neural networks, which includes a fixed target network $T$ and a learnable predictor network $P$.
Both networks have identical neural architectures, consisting of a simple multi-layer perceptron (MLP)~\cite{haykin1994neural}. The fixed target network is initialized with random weights, which remain unchanged throughout the fuzzing process. In contrast, the predictor network is trained to approximate the output of the target network. 
For each encountered state, we denote the output generated by the target network as $T(s)$ and the output generated by the predictor network as $P(s)$. The difference between the outputs of these two networks, referred to as the prediction error, serves as a proxy for the novelty of a given state. The prediction error, essentially is the mean squared error (MSE) between the outputs of the target and predictor networks. It is computed as:
\begin{equation}
\label{equation:rnd}
  E(s) = ||T(s) - P(s)||^2
\end{equation}

This prediction error serves as an intrinsic reward signal for each state. In \toolname, given the induced states sequence of a scenario, we calculate the intrinsic reward for all states within the sequence. The novelty score of a scenario is the mean of all intrinsic rewards. 
The intrinsic reward for the `novel' inputs should be higher than the previously encountered inputs during the fuzzing process, thereby driving the fuzzer to explore those novel states. In addition, when updating the parameters of the predictor network, we apply $L_2$ regularization to avoid overfitting.

The quantity of training data will influence the magnitude of prediction errors, which explains why our curiosity mechanism serves as a reliable novelty measure for scenarios. 
When the predictor network encounters few instances for certain types of scenario during training, the prediction error tends to be high. 
Conversely, when the SDM frequently encounters one type of scenario, the predictor network has more opportunities to learn and mimic the target network's responses for that specific scenario. 
The predictor network can accurately predict the outcomes and the prediction error tends to be lower, indicating these are familiar scenarios. 
By leveraging the prediction error as a measure of curiosity, our mechanism can effectively identify scenarios that deviate from the learned patterns, highlighting their novelty. 
This approach enables the fuzzer to prioritize exploring unfamiliar scenarios, leading to an increased diversity of crash-triggering scenarios during fuzz testing for SDMs.


\subsection{\toolname Architecture}
\label{subsec:architecture}
\begin{algorithm}[!t]
  \caption{\toolname Workflow}
  \label{algo:workflow}

      \SetKwInOut{Input}{Input}
      \SetKwInOut{Output}{Output}
      \Input{Target SDM: $SDM$, Environment: $env$}
      \Output{Crash-triggering Seeds: $C$}
      \SetKwFunction{InitCorpus}{InitCorpus}
      \SetKwFunction{EnvSimulator}{EnvSim}
      \SetKwFunction{EnvMonitor}{EnvMonitor}
      \SetKwFunction{SeedSelection}{SeedSelection}
      \SetKwFunction{SeedMutation}{SeedMutation}
      \SetKwFunction{isCrashed}{isCrashed}
      \SetKwFunction{isInteresting}{isInteresting}
      \SetKwFunction{initCuriosity}{initCuriosity}
      \SetKwFunction{Curiosity}{Curiosity}
      \SetKwFunction{CureFuzz}{CureFuzz}
      \SetKwProg{Fn}{Function}{:}{}

      \Fn{\EnvMonitor{$env, SDM, s$}}{
        $ \{s'\}, r, \longleftarrow \EnvSimulator(env, SDM, s, max\_step)$\;
        \KwRet $\{s'\}, r $\;
      }

      \Fn{\InitCorpus{}}{
      $I \longleftarrow \emptyset$\ 

      \While{ runnign time $<$ 2 hours}{
        $s \longleftarrow RandomState(env)$\;
        $s' \longleftarrow \EnvMonitor($env, SDM, s$)$\;

        $I \longleftarrow I \cup \{s'\}$\;
      }
      \KwRet $I$\;
      }

      \Fn{\CureFuzz{$env, SDM$}}{
        $C \longleftarrow \emptyset$\;
        $\theta^T, \theta^P \longleftarrow \initCuriosity$\;
        $I \longleftarrow \InitCorpus($env, SDM$)$\;

      \While{ runnign time $<$ 12 hours}{
          $s \longleftarrow \SeedSelection(I)$\;
          $s_{\delta} \longleftarrow \SeedMutation(s)$\;
          $s'_{\delta} \longleftarrow \EnvMonitor(env, SDM, s_{\delta})$\;
          $in\_reward, \theta^T, \theta^P \longleftarrow \Curiosity( s'_{\delta}, \theta^T, \theta^P)$\;
          \uIf{ \isCrashed ($s'_{\delta}$)}{
            $C \longleftarrow C \cup s'_{\delta}$\;
          }
          \uElseIf{\isInteresting ($s'_{\delta}$)}{
            $I \longleftarrow I \cup s'_{\delta}$\;
          }
      }
      \KwRet $C$\;
}
\end{algorithm}

\vspace*{0.2cm}
\noindent \textbf{Setup and Initialization}: Lines 13-15 in Algorithm \ref{algo:workflow} shows the initialization of \toolname. The function \textit{EnvMonitor} is responsible for monitoring the interaction between the SDM and the environment. 
For the target SDM and environment, given an initial state as the input, \textit{EnvMonitor} returns the induced state sequence and the associated cumulative reward. \toolname first loads the target SDM and the environment. $\theta^T$ and $\theta^P$ are the parameters of our curiosity module and are randomly initialized. The fuzzer then generates an initial corpus of seed by randomly sampling with the environment, as depicted in Function \textit{InitCorpus} (Algorithm \ref{algo:workflow} Line 5 to 11). The underlying assumption is that we are aware of the legitimate state space of the environment, allowing us to generate valid seeds across this state space randomly. Since the definition of legitimate state space varies in different environments, a detailed description for each environment is given in Section \ref{subsec:env}. 


\vspace*{0.2cm}
\noindent \textbf{Seed energy estimation}: The main fuzzing process starts at Line 15. Following the settings from prior works~\cite{MDPFuzz,KleesRCW018}, a 12-hour long fuzzing is conducted.
At each iteration, a seed is selected from the seed corpus. Similar to traditional fuzz testing of software~\cite{takanen2018fuzzing}, we first estimate the energy of each seed. A seed with a high energy is more likely to be selected. Because \toolname aims to find a diverse set of crashes, the energy of a seed cannot only reflect the novelty but also needs to consider the probability of triggering crashes. To balance these objectives, the estimation process is based on multiple factors. These factors are intrinsic reward (novelty measure), cumulative reward (probability of triggering crashes), and robustness (probability of triggering unseen states).

\noindent\textit{Cumulative reward}. The cumulative reward is a direct measurement on the performance of the SDM. We begin with a simple assumption: if the SDM does not perform well under certain scenarios, mutating on these scenarios is more likely to trigger crashes. \toolname prioritizes the seeds with low cumulative reward. Thus a seed with a lower cumulative reward is accordingly given a higher energy estimation.

\noindent\textit{Robustness}.
We define the term Robustness as the probability for a given seed to trigger diverse consequences. 
The calculation of robustness is detailed in the Algorithm~\ref{algo:robust}.
Given a seed $s$ and its induced state sequence $\boldsymbol{S}$, we first record the final state of $\boldsymbol{S}$, denoted as $\boldsymbol{S}_{-1}$.
We add a tiny random perturbation on $s$ to generate a new initial state $s^{\delta}$. 
The mutated state $s^{\delta}$ then is fed the function \textit{EnvMonitor}, and we obtain its induced sequence $\boldsymbol{S}^{\delta}$. 
Robustness is then measured as the Euclidean distance between the final states of $\boldsymbol{S}$ and $\boldsymbol{S}^{\delta}$, which essentially is $ |\boldsymbol{S}_{-1} - \boldsymbol{S}^{\delta}_{-1}|$. 

Robustness is a measure of how sensitive an SDM's behavior is to slight perturbations in the original state. When the Euclidean distance between the final states of the original and mutated state sequences is large, it indicates that even a small perturbation in the initial state leads to significantly different outcomes. This suggests that the seed has the potential to trigger a diverse range of behaviors, making it a valuable candidate for further exploration and testing. On the other hand, if the distance between the final states is small, it implies that the SDM's behavior is relatively consistent and less sensitive to small changes in the input. In this case, the seed might be less likely to reveal new or unexpected system behaviors.

\noindent\textit{Intrinsic reward}.
The intrinsic reward generated by the curiosity mechanism serves as the novelty measure of a given seed. The intrinsic reward can be easily combined with the cumulative reward and robustness score measurement.

\begin{algorithm}[!t]
  \DontPrintSemicolon
  \caption{Robustness Estimation}
  \label{algo:robust}
  \SetKwInOut{Input}{Input}
        \SetKwInOut{Output}{Output}
        \Input{State sequence: $\boldsymbol{S}, env, SDM $}
        \Output{Robustness Measure: $R$}
        \SetKwFunction{Robustness}{Robustness}
        \SetKwFunction{SeedMutation}{SeedMutation}
        \SetKwFunction{EnvMonitor}{EnvMonitor}

        \SetKwProg{Fn}{Function}{:}{}

        \Fn{\Robustness{$\boldsymbol{S}$}}{
          Obtain $ s_{0}$ from $ \boldsymbol{S} $\;
          $ s^{\delta}_0 \longleftarrow \SeedMutation(s_0)$\;
          $\boldsymbol{S}^{\delta}, r\longleftarrow $ \EnvMonitor($env, SDM, s^{\delta}_0,$)

          \KwRet $ |\boldsymbol{S}_{-1} - \boldsymbol{S}^{\delta}_{-1}|$\;

        } 
  \end{algorithm}  
  Denoting reward as $r$, intrinsic reward as $i$, robustness as $r'$, $\alpha, \beta, \text{and } \gamma$ as scaling factors, the overall score of seed is:
\begin{equation}
  E(s) = e^{-\alpha r}+e^{\beta i}+ \gamma r'
\end{equation}

\noindent \textbf{Seed Selection}:
We prioritize selecting the seed with a high energy. Given the corpus $C$ and total number of seeds in the corpus $N$, each seed would be selected with a probability of $\frac{ E(s)}{\sum_{i=1}^{N} E(s_i)}$, where $\sum_{i=1}^{N} E(s_i)$ denotes the total energy of the seeds in the corpus.

\noindent\textbf{Seed Mutation}: 
Once a seed is selected from the corpus, the \textit{SeedMutation} function generates a new mutated state by applying a small random perturbation to a selected state. The mutated state is fed into the \textit{EnvMonitor} function to generate its corresponding state sequence and collect the cumulative reward. We need to ensure that the mutated seed lies in the legitimate state spaces 
, and the mutated seed would not trigger an initial crash. In our experiments, we have carefully addressed this concern and verified the validity of each mutated seed, details are given in Section~\ref{sec:exp}. 

\noindent\textbf{Seed Evaluation}:
The function \textit{Curiosity} assigns the intrinsic reward to the newly induced state sequence, and this intrinsic reward serves as the curiosity of the fuzzer in the further mutation of this seed. For each state sequence, we calculate the difference between the two networks of the curiosity mechanism using MSE loss. The parameters of the predictor network are updated using this loss. 
\toolname then checks on the termination status of the state sequence. If a crash is found, the mutated seed is added to the list of crashes, and the fuzzing process continues with the next seed input. 
When \toolname does not identify a crash, it shifts focus to assessing the state sequence based on the induced intrinsic reward and cumulative reward.  \toolname measures the intrinsic reward of the state sequence. If this reward surpasses a pre-set threshold, the seed responsible for this sequence is considered significant. Consequently, it is added to the seed corpus for further analysis.
\toolname also compares the cumulative reward of the mutated seed against that of the original seed. This mutated seed is also added to the corpus if the mutated seed's cumulative reward is lower. 
If neither of these conditions are met, the mutated seed is then discarded, and \toolname progresses to the next iteration.

\vspace*{0.2cm}
\noindent\textbf{Termination}: The fuzzing process continues the above-mentioned iteration and is terminated when the time has exceeded the specified limit (12 hours). Finally, \toolname returns the list of crashes found during the fuzzing process.

\section{Experimental Setting}
\label{sec:exp}
\subsection{Research Questions}
To evaluate the performance of \toolname and comprehensively understand the impact, we formulate the three research questions:

\noindent \textbf{RQ1: How effective is \toolname in finding crash-triggering scenarios?}


\noindent \textbf{RQ2: Can \toolname be effectively guided with the curiosity mechanism?}

\noindent \textbf{RQ3: Can we use the crashes found by \toolname to improve SDMs?}




\label{subsec:env}
\subsection{Experiment Subject and Environment}

We evaluate \toolname using the same environments and SDMs
as Pang et al.~\cite{MDPFuzz}. 
Our investigation covers various environments, including the CARLA autonomous driving simulator~\cite{DosovitskiyRCLK17}, ACAS Xu for collision avoidance in aviation~\cite{marston2015acas}, the Cooperative Navigation (Coop Navi) environment for multi-agent reinforcement learning~\cite{LoweWTHAM17}, and the BipedalWalker environment in OpenAI Gym~\cite{kuznetsov2020controlling}. 

\subsubsection{\textbf{Autonomous Driving}}
CARLA~\cite{DosovitskiyRCLK17}
is a widely used open-source simulator for autonomous driving research.
In the CARLA simulator, at each timestep, the SDM receives an RGB image and its current velocity as inputs. Using this information, the SDM calculates the appropriate steering, throttle, and brake commands to navigate toward the specified goals. The performance of the SDM is assessed in an urban driving environment that includes intersections and traffic lights. 
Two SDMs are considered in the CARLA environment, which are developed using DRL and IL, respectively. 
In CARLA, \toolname checks for the situations when the SDM-controlled vehicle experiences a collision with other vehicles or buildings. The environment of CARLA can be described as the positions of angles of all vehicles on the map, including the SDM-controlled one. When \toolname mutates a given state, small perturbations are randomly generated and added to the positions and angles of these vehicles. We use the CARLA simulator itself to check for the validity of the mutated state. All illegally mutated states and the states that trigger initial collision are discarded in our experiments.

\subsubsection{\textbf{Aircraft Collision Avoidance}}
ACAS Xu is a collision avoidance system for the aviation industry~\cite{marston2015acas}. We utilize the popular DNN-based variant of ACAS Xu~\cite{acasxu}, which has also been broadly studied by previous literature~\cite{WangPWYJ18}. The system employs 45 distinct neural networks to predict the most appropriate actions to avoid the collision, such as 
Clear-of-Conflict (which goes straight), weak left (1.5 deg/s), strong left (3.0 deg/s), weak right, and strong right. In ACAS Xu, \toolname simply aims to find the scenarios when there are collisions between the SDM-controlled airplane with other airplanes. For mutation, the initial positions and speeds of the SDM-controlled and the other planes slightly changed. The maximal speed of all airplanes is capped at 1,100 ft/sec. 
Given the predefined range of acceptable states, states that fall outside legal space are automatically discarded.

\subsubsection{\textbf{Video Game}}
Coop Navi~\cite{LoweWTHAM17} is an OpenAI environment designed for multi-agent reinforcement learning (MARL) applications. In Coop Navi, agents aim to learn how to cooperate with each other to reach the pre-determined landmarks without colliding with one another. The underlying SDM for this game is proposed by the original publication~\cite{LoweWTHAM17}. This SDM is aware of the position of each agent and the target landmarks, then the SDM decides the moving direction and speed for each agent. \toolname aims to find the scenarios when there are collisions between SDM-controlled agents. The initial positions of the three MARL-controlled agents are mutated, and their initial speeds are set to 0 to avoid initial collision. With a clear definition of permissible positions and velocities, \toolname ignores any illegal states.

BipedalWalker~\cite{kuznetsov2020controlling} is an environment in the OpenAI Gym framework that challenges a two-legged robot to navigate through various terrains and obstacles using bipedal locomotion. We select the SDM which is employed with Twin Delayed DDPG with Quantile Distributional Critics (TQC)~\cite{kuznetsov2020controlling} algorithm, and its implementation is available in the stable-baselines3 repository~\cite{rlbaselines3}. The robot takes in a 24-dimensional state and predicts the speed for each leg based on body angle, leg angles, speed, and lidar data.
We aim to find the scenarios when the robot falls. Following Pang et al.~\cite{MDPFuzz}, we mutate the sequence of ground types the robot encounters. Specifically, we make sure that the first 20 frames are ``flat'' to prevent initial failure. We also place a ``flat'' between two hurdles to enable the agent to pass the obstacles while taking optimal actions. Since the valid ground types are known, illegal states can be easily detected and discarded.

\subsection{Implementation}

\toolname is coded in Python and the curiosity module is implemented with the Pytorch Library~\cite{NEURIPS2019_9015}. The curiosity module utilizes a simple multilayer perceptron~\cite{haykin1994neural} as the underlying neural architecture for both the target network and predictor network, and we use the ReLU function~\cite{agarap2018deep} as the activation function. 
Following prior work~\cite{MDPFuzz}, we randomly sample for 2 hours to construct the initial corpus, and the main fuzzing process is conducted for 12 hours as the standard setting~\cite{KleesRCW018}. 
The interaction time between the SDMs and the environment of Coop Navi and Acas Xu takes significantly less time than the other environments. We follow Pang et al.~\cite{MDPFuzz} to modify the experiment setup accordingly. Specifically, we reduce the time taken to construct the initial corpus to 1 hour in ACAS Xu, and 30 minutes in Coop Navi, then conduct the 12 hours of fuzzing. For each SDM, we make sure \toolname and MDPFuzz are experimented under the same setting to make a fair comparison. 
When mutating a seed, the magnitude of the random perturbation is a critical factor that can impact the performance of the fuzzer. We re-use the code implementation from Pang et al. for generating mutations. 
Our experiments are conducted on a server with one Intel Xeon E5-2698 v4 @ 2.20GHz CPU, 504GB RAM, and NVIDIA Tesla V100 GPU.

\section{Experiment Results}
\label{sec:result}

\subsection*{\textbf{RQ1: How effective is \toolname in finding
crash-triggering scenarios?}}

\noindent We compare the performance of \toolname with two state-of-the-art approaches: MDPFuzz~\cite{MDPFuzz} and the method proposed by Li et al.~\cite{diffusion} (referred to as ``G-Model'' in our paper). MDPFuzz is a black-box fuzz testing framework for SDMs while G-Model employs a model-based method to generate diverse scenarios. 
Our evaluation focuses on three key metrics: environmental state coverage, the total number of detected crash-triggering scenarios, and the distinct types of crash-triggering scenarios.  
We repeat the experiment five times for each SDM and report the average results. The inclusion of two markedly different baselines and three diverse metrics ensures that our evaluation comprehensively compares the effectiveness of our \toolname and the baselines from multiple perspectives.

\noindent\textsf{\emph{Coverage Analysis.}} Coverage analysis plays a crucial role in evaluating the thoroughness and completeness of a testing technique. Inspired by fuzzing for traditional software, which measures code coverage, we measure the environmental state coverage in our experiments. State coverage refers to the effectiveness of a fuzzer in covering the possible scenarios within the state space of the environment. Considering that experimental environments have a high-dimensional and continuous state space, to calculate state coverage, we discretize the state space of the experimental environments by bucketing continuous variables into discrete segments. In other words, we divide each dimension of the state space into a fixed number of bins, effectively creating a grid over the state space. Each bin represents a discrete state. When performing the state discretization, we use multiple numbers of bins (i.e., 5, 10, 100)~\cite{lopez2018demand}. Note that the exception here is the case of BipedalWalker, where we mutate the sequence of ground types that the robot needs to walk through. Since the available number of ground types is fixed, we only need to calculate the proportion of encountered ground types over the total number of ground types as state coverage and there are no actual bins. We present the average state coverage results of \toolname and the baselines in Table~\ref{tab:coverage}.  \toolname demonstrates promising state coverage performance in comparison to baseline methods across various environments and bin numbers. We perform the Mann-Whitney U~\cite{nachar2008mann} statistical significance test at 95\% significance level and compute effect size measure (Cohen’s d~\cite{grissom2005effect}). Notable environments where it significantly and substantially outperforms others include Carla(RL), Acas Xu (DNN), Coop Navi(MARL), and BipedalWalker(RL). The only exception is in Carla(IL), where \toolname shows comparable results to G-Model. Overall, \toolname achieves the best coverage rate.

\begin{figure}[!t]
  \centering
  \includegraphics[width=0.8\linewidth]{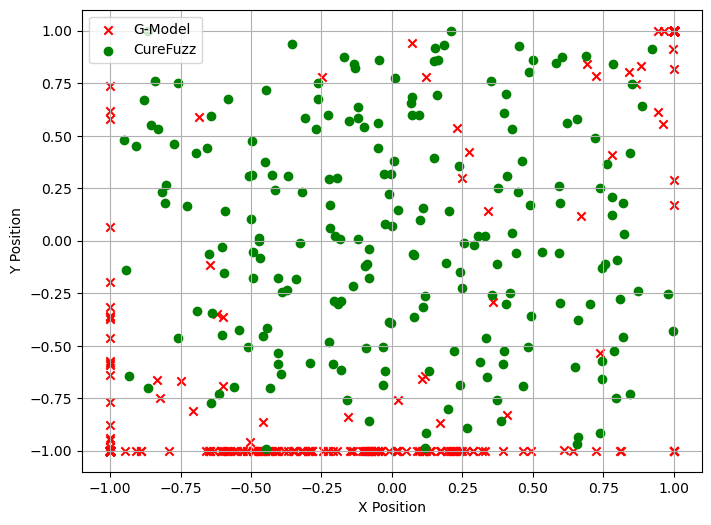}
  \caption{Visualization of Agent Positions in Crash-Triggering Scenarios found by various methods for Coop Navi (MARL). This graph plots the x and y coordinates, ranging from -1 to 1. The red crosses represent G-model, the green dots represent \toolname, and the blue triangles represent MDPFuzz. }
  \label{fig:output}
\end{figure}

\begin{table}[]
\centering
\caption{State Coverage of \toolname and baselines with 95\% confidence interval margins of error.}
\label{tab:coverage}
\resizebox{\columnwidth}{!}{%
\begin{tabular}{cl|c|c|c}
\hline
                                             & \textbf{Method}   & \textbf{5 bins} & \textbf{10 bins} & \textbf{100 bins} \\ \hline
\multirow{3}{*}{\textbf{Carla (RL)}}         & \textbf{MDPFuzz}  & 7.3\% $\pm$ 0.6\% & 1.3\% $\pm$ 0.1\% & $3 \mathrm{e}^{-5}$ $\pm$ $3 \mathrm{e}^{-6}$ \\
                                             & \textbf{G-Model}  & 5.0\% $\pm$ 1.2\% & 1.6\% $\pm$ 0.3\% & $2.4 \mathrm{e}^{-5}$ $\pm$ $5.5 \mathrm{e}^{-6}$ \\
                                             & \textbf{CureFuzz} & 8.0\% $\pm$ 0.9\% & 3.4\% $\pm$ 0.3\% & $1 \mathrm{e}^{-4}$ $\pm$ $2 \mathrm{e}^{-5}$   \\ \hline
\multirow{3}{*}{\textbf{ACAS Xu (DNN)}}      & \textbf{MDPFuzz}  & 5.3\% $\pm$ 0.5\% & 0.48\% $\pm$ 0.01\% & $2 \mathrm{e}^{-6}$ $\pm$ $4 \mathrm{e}^{-8}$ \\
                                             & \textbf{G-Model}  & 6.7\% $\pm$ 0.9\% & 0.72\% $\pm$ 0.02\% & $3 \mathrm{e}^{-6}$ $\pm$ $1 \mathrm{e}^{-7}$ \\
                                             & \textbf{CureFuzz} & 11\% $\pm$ 0.6\% & 1.2\% $\pm$ 0.15\% & $6 \mathrm{e}^{-6}$ $\pm$ $1 \mathrm{e}^{-6}$   \\ \hline
\multirow{3}{*}{\textbf{Carla(IL)}}         & \textbf{MDPFuzz}  & 6.4\% $\pm$ 1.2\% & 1.0\% $\pm$ 0.6\% & $2 \mathrm{e}^{-5}$ $\pm$ $2 \mathrm{e}^{-6}$ \\
                                             & \textbf{G-Model}  & 6.7\% $\pm$ 0.5\% & 1.6\% $\pm$ 0.5\% & $3 \mathrm{e}^{-5}$ $\pm$ $1 \mathrm{e}^{-6}$ \\
                                             & \textbf{CureFuzz} & 7.2\% $\pm$ 0.1\% & 1.6\% $\pm$ 0.44\% & $3 \mathrm{e}^{-5}$ $\pm$ $1 \mathrm{e}^{-7}$  \\ \hline
\multirow{3}{*}{\textbf{Coop Navi (MARL)}  }   & \textbf{MDPFuzz}  & 0.034\% $\pm$ 0.008\% & $3 \mathrm{e}^{-7}$ $\pm$ $1 \mathrm{e}^{-8}$ & $5 \mathrm{e}^{-19}$ $\pm$ $6 \mathrm{e}^{-20}$ \\
                                             & \textbf{G-Model}  & 0.021\% $\pm$ 0.008\% & $6 \mathrm{e}^{-8}$ $\pm$ $3 \mathrm{e}^{-9}$ & $7 \mathrm{e}^{-20}$ $\pm$ $3 \mathrm{e}^{-21}$ \\
                                           & \textbf{CureFuzz} & 0.059\% $\pm$ 0.018\% & $7 \mathrm{e}^{-7}$ $\pm$ $1 \mathrm{e}^{-7}$ & $1 \mathrm{e}^{-18}$ $\pm$ $5 \mathrm{e}^{-19}$ \\ \hline
\multirow{3}{*}{\textbf{BipedalWalker (DRL)}}                  & \textbf{MDPFuzz}  & 0.065\% $\pm$ 0.002\% & {--} & {--} \\
                                              & \textbf{G-Model}  & 0.011\% $\pm$ 0.002\% & {--} & {--} \\
                                             & \textbf{CureFuzz} & 0.42\% $\pm$ 0.02\% & {--} & {--} \\ \hline
\end{tabular}%
}
\end{table}
\noindent\textsf{\emph{Total Number of Crashes.}}
Table~\ref{tab:rq1-result} shows the comparative results of \toolname and baseline methods in terms of the number of detected crashes. The results show that \toolname consistently outperforms MDPFuzz in all five sets of experiments. The most significant improvement is observed in the SDM of BipedalWalker(RL), with a remarkable 422.22\% increase in detected crashes, rising from 126 to 658. For CARLA(RL) and CARLA(IL), the improvements are 145.83\% (from 120 to 295) and 110.47\% (from 86 to 181), respectively; for Coop Navi(MARL), the performance is boosted by 62.21\% (from 52.4 to 85). The smallest improvement is for ACAS Xu(DNN), which is 46.45\% (from 183 to 268). 
Furthermore, \toolname outperforms G-Model for four out of five SDMs by up to 134.16\% (from 281 to 658 in BipedalWalker(RL)). For Carla(RL) and Carla(IL), the improvements are 86.7\% (from 158 to 195)and 60.1\% (from 113 to 181). \toolname also archives an improvement of 94.2\% in Acas Xu(DNN).
After performing the statistical test and computing effect size, results show that CureFuzz statistically significantly and substantially outperforms both baselines in these cases. 
The only exception is in the context of Coop Navi(MARL), where G-Model identifies more crashes than \toolname (185.4 vs. 85). We perform additional analysis to investigate why it is happening for Coop Navi(MARL). Recall that the goal of the Coop Navi game is to find the scenarios when the agents collide with each other, in Figure \ref{fig:output}, we select the median performance from the five repeated experiments for \toolname and G-Model, and plot the positions of the agents in the crash-triggering scenarios found by different methods. While the valid range of the agent's position is from -1 to 1, from Figure \ref{fig:output}, we can see that G-Model tends to find the crashes that the agents are positioned around the boundary of the valid input. In contrast, \toolname covers more spread positions. It also explains why the coverage of G-Model is lower than \toolname but it finds more crashes for Coop Navi (MARL). \toolname and G-Model tend to be interested in a different area of the search space, thus one possible future direction could combine the advantages of both kinds of methods. 

\noindent\textsf{\emph{Distinct Crashes.}}
Using the same state discretization procedure that we used for coverage analysis, we transform the continuous state space into a discrete grid representation again. For each method, we determine the average number of unique grid cells occupied by the detected crashes. We account for one grid cell as one distinct crash. As shown in Table~\ref{tab:distinct-table}, \toolname finds more types of crash-triggering scenarios than MDPFuzz in all bin numbers and every SDM. For the 100 bins situation, the largest improvement is seen for the Carla(IL) SDM, where \toolname found 200\% more types than MDPFuzz. Similarly, for the other SDMs, \toolname found 88.9\%, 177.8\%, and 62.8\% more types of crashes respectively. After performing the statistical test and computing effect size, we find that CureFuzz statistically significantly and substantially outperforms the baselines apart from Coop Navi (MARL). Again, in the context of Coop Navi (MARL), \toolname falls short of G-Model. We have discussed this in detail in the sub-section above.

\begin{table}[]
\centering
\caption{Average total number of crashes found by \toolname and the baselines run across different SDMs and margin of errors in 95\% confidence interval.}
\footnotesize
\label{tab:rq1-result}
\begin{tabular}{llc}
\toprule
                                        & \textbf{Method}     & \textbf{Mean} \\
\midrule
\multirow{3}{*}{\textbf{Carla (RL)}}         & \textbf{MDPFuzz}    & 120 $\pm$ 22.8 \\
                                            & \textbf{G-Model}    & 158 $\pm$ 18.6 \\
                                            & \textbf{CureFuzz}   & 295 $\pm$ 38.3 \\
\midrule
\multirow{3}{*}{\textbf{ACAS Xu (DNN)}}      & \textbf{MDPFuzz}    & 183 $\pm$ 15.9 \\
                                            & \textbf{G-Model}    & 138 $\pm$ 27.9 \\
                                            & \textbf{CureFuzz}   & 268 $\pm$ 26.8 \\
\midrule
\multirow{3}{*}{\textbf{Carla (IL)}}         & \textbf{MDPFuzz}    & 86 $\pm$ 25.0  \\
                                            & \textbf{G-Model}    & 113 $\pm$ 33.5 \\
                                            & \textbf{CureFuzz}   & 181 $\pm$ 25.4 \\
\midrule
\multirow{3}{*}{\textbf{Coop Navi (MARL)}}   & \textbf{MDPFuzz}    & 52.4 $\pm$ 8.8 \\
                                            & \textbf{G-Model}    & 185.4 $\pm$ 36.6 \\
                                            & \textbf{CureFuzz}   & 85 $\pm$ 7.3   \\
\midrule
\multirow{3}{*}{\textbf{BipedalWalker (RL)}} & \textbf{MDPFuzz}    & 126 $\pm$ 31.8  \\
                                            & \textbf{G-Model}    & 281 $\pm$ 52.5  \\
                                            & \textbf{CureFuzz}   & 658 $\pm$ 98.3  \\
\bottomrule
\end{tabular}
\end{table}

\begin{table}[]
\centering
\caption{Average distinct types of crash-triggering scenarios found by \toolname and baselines run across different SDMs and margin of errors in 95\% confidence interval.}
\label{tab:distinct-table}
\resizebox{\columnwidth}{!}{%
\begin{tabular}{cl|c|c|c}
\hline
                                             & \textbf{Method}   & \textbf{5 bins}     & \textbf{10 bins}    & \textbf{100 bins}   \\ \hline
\multirow{3}{*}{\textbf{Carla (RL)}}         & \textbf{MDPFuzz}  & 9.0 $\pm$ 0.9     & 17.8 $\pm$ 2.4    & 29.8 $\pm$ 3.9    \\
                                             & \textbf{G-Model}  & 6.8 $\pm$ 1.0     & 12.2 $\pm$ 1.1     & 20.6 $\pm$ 1.5    \\
                                             & \textbf{CureFuzz} & 10.0 $\pm$ 0.6    & 30.2 $\pm$ 3.1    & 89.4 $\pm$ 25.3    \\ \hline
\multirow{3}{*}{\textbf{ACAS Xu (DNN)}}      & \textbf{MDPFuzz}  & 8.0 $\pm$ 0.9     & 9.0 $\pm$ 0.9     & 9.0 $\pm$ 0.9     \\
                                             & \textbf{G-Model}  & 5.8 $\pm$ 1.6       & 6.4 $\pm$ 1.1     & 6.6 $\pm$ 1.7     \\
                                             & \textbf{CureFuzz} & 12.0 $\pm$ 0.9    & 13.6 $\pm$ 0.7    & 17.0 $\pm$ 1.5    \\ \hline
\multirow{3}{*}{\textbf{Carla (IL)}}         & \textbf{MDPFuzz}  & 9.0 $\pm$ 0.9     & 14.4 $\pm$ 2.3    & 21.6 $\pm$ 3.0    \\
                                             & \textbf{G-Model}  & 5.6 $\pm$ 0.7     & 12.8 $\pm$ 3.2    & 23.4 $\pm$ 6.0    \\
                                             & \textbf{CureFuzz} & 10.0 $\pm$ 0.0    & 27.5 $\pm$ 1.9    & 60.0 $\pm$ 10.5    \\ \hline
\multirow{3}{*}{\textbf{Coop Navi (MARL)}}   & \textbf{MDPFuzz}  & 52 $\pm$ 11.0      & 52.4 $\pm$ 11.7    & 52.4 $\pm$ 11.7    \\
                                             & \textbf{G-Model}  & 104.8 $\pm$ 35.5  & 156.2 $\pm$ 30.0  & 184.6 $\pm$ 38.8  \\
                                             & \textbf{CureFuzz} & 83.5 $\pm$ 8.0   & 85.3 $\pm$ 7.3   & 85.3 $\pm$ 7.3   \\ \hline
\multirow{3}{*}{\textbf{BipedalWalker (RL)}} & \textbf{MDPFuzz}  & 126 $\pm$ 31.8    & -                  & -                  \\
                                             & \textbf{G-Model}  & 281 $\pm$ 52.5  & -                  & -                  \\
                                             & \textbf{CureFuzz} & 658 $\pm$ 98.3    & -                  & -                  \\ \hline
\end{tabular}%
}
\end{table}

\vspace{0.1cm}
\noindent\textsf{\emph{Efficiency Analysis.}}
We further conduct an in-depth analysis to investigate the efficiency of \toolname. The total time taken by a fuzz testing method can be split into two parts, which are execution time and analysis time. The execution time corresponds to the duration an SDM needs to interact with its environment, while the analysis time accounts for the computational time needed for the fuzzing process (i.e., seed selection and seed evaluation). Table~\ref{tab:time} presents the average analysis time for each iteration of \toolname and MDPFuzz. Note here we do not compare with G-Model in the efficiency analysis since it operates differently. It is not a fuzzing test method and requires the retraining of its generative model in each iteration. This retraining process can significantly extend the time taken, making it incomparable to the efficiency of \toolname and MDPFuzz, as these two fuzzing methods do not necessitate repeated training. The results clearly demonstrate the remarkable efficiency of \toolname, with its analysis time typically requiring only up to 0.08 seconds. In contrast, MDPFuzz takes about one second in some cases. On average, \toolname's analysis time is 81.8\% faster than that of MDPFuzz, highlighting the minimal time overhead associated with our curiosity mechanism across various environments and state spaces.

\begin{table}[!t]
    \centering
    \caption{Time comparison between CureFuzz and MDPFuzz in seconds.}
    \label{tab:time}
    \begin{tabular}{ccc}
    \toprule
    \textbf{} & \textbf{MDPFuzz} & \textbf{CureFuzz} \\
    \midrule
    Carla (RL) & 0.855 & 0.008 \textbf{(+99.1\%)} \\
    ACAS Xu (DNN) & 0.231 & 0.005 \textbf{(+97.8\%)}\\
    Carla (IL) & 0.985 & 0.005 \textbf{(+99.5\%)} \\
    Coop Navi (MARL) & 0.033 & 0.006 \textbf{(+81.8\%)} \\
    BipedalWalker (RL) & 0.970 & 0.011 \textbf{(+98.9\%)}\\
    \bottomrule
    \end{tabular}
\end{table}
\vspace{5pt}
\ans{
    \textbf{Answer to RQ1}: \toolname can efficiently find crash-triggering scenarios across various SDMs and environments. 
    \vspace{5pt}

}

\subsection*{\textbf{RQ2: Can \toolname be effectively guided with the curiosity mechanism?}}

\begin{figure*}[!t]
    \centering
    \subfloat[ACAS Xu(DNN) \label{fig2:acas}]{\includegraphics[width = 0.2\linewidth]{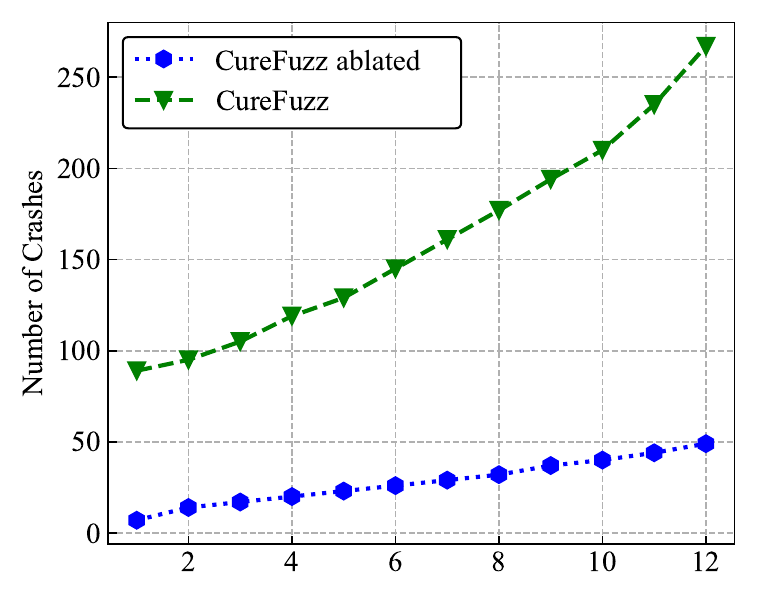}}
    \hfil
    \subfloat[Carla(IL) \label{fig2:ilcarla}]{\includegraphics[width = 0.2\linewidth]{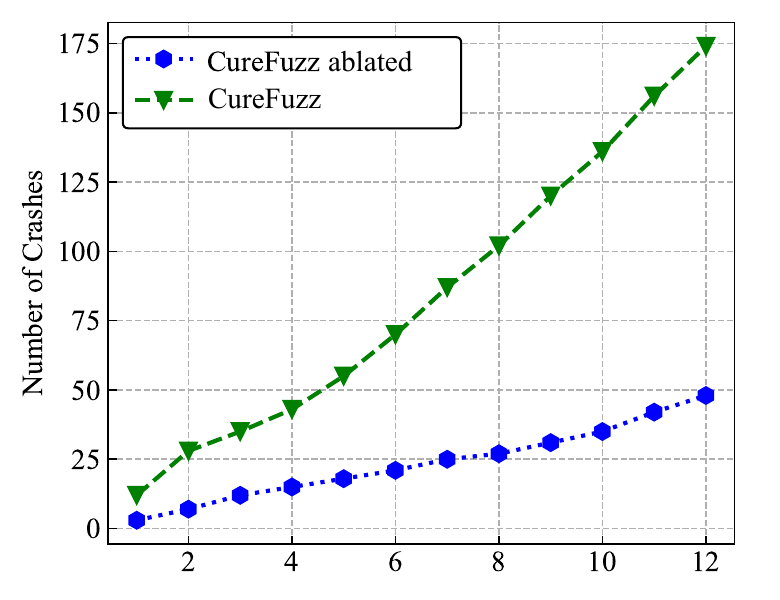}}
    \hfil
    \subfloat[Carla(RL) \label{fig2:rlcarla}]{\includegraphics[width = 0.2\linewidth]{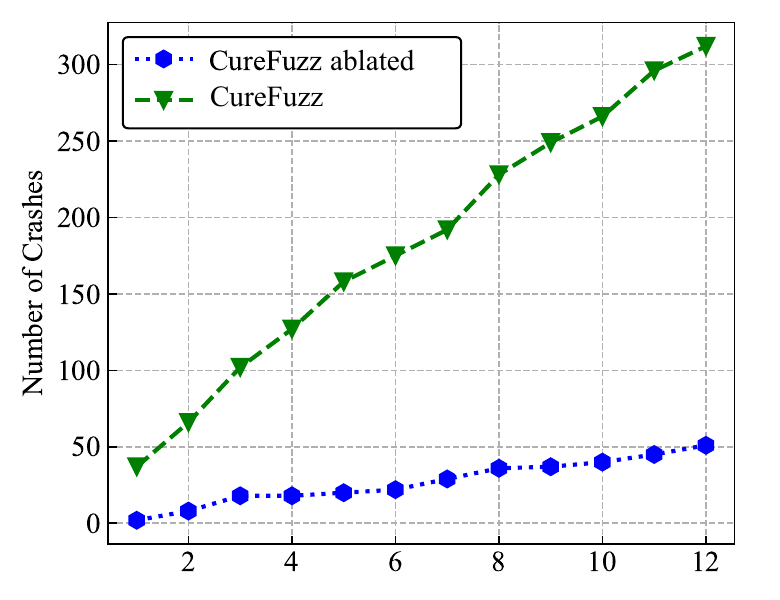}}
    \hfil
    \subfloat[Coop Navi(MARL) \label{fig2:marl}]{\includegraphics[width = 0.2\linewidth]{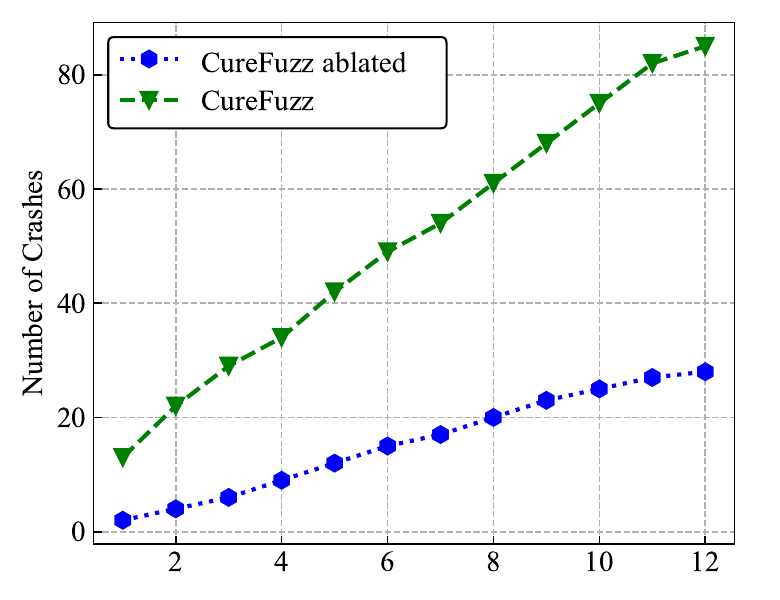}}
    \hfil
    \subfloat[BipedalWalker(RL) \label{fig2:walker}]{\includegraphics[width = 0.2\linewidth]{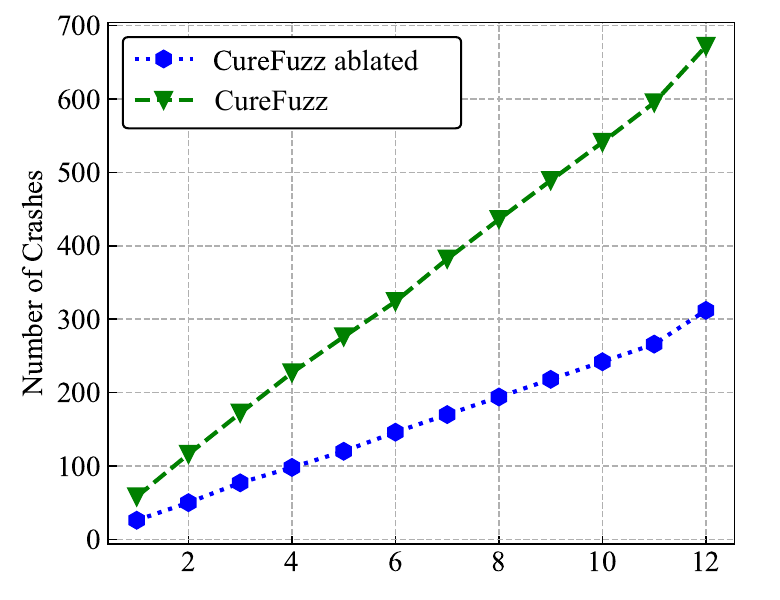}}
    \caption{Results for RQ2: Comparison between \toolname with and without the curiosity mechanism. The x-axis represents the time passed in hours and the y-axis represents the number of found crashes. The complete \toolname is represented with green color and triangles, the ablated \toolname is represented with blue color and circles. }
    \label{fig:rq2}
    \end{figure*}
For this research question, we follow the same experimental setting with RQ1. 
We remove the curiosity component from \toolname and conduct the fuzzing again on our target SDMs. 
Table \ref{tab:rq2} shows the performance of the ablated \tool. We can observe that the addition of the curiosity mechanism plays a vital role in boosting the effectiveness of \toolname, as it provides an additional signal that motivates \toolname to further explore the states it is curious about. 
We plot the number of crashes found by \toolname and the ablated \toolname on an hourly basis in Figure~\ref{fig:rq2}. Notice that the experiments are repeated five times, and we select the median performer for plotting. We can see that \toolname consistently outperforms ablated \toolname across all scenarios and time intervals.

\begin{table}[]
\centering
\caption{Results for RQ2, shows the average state coverage (for 100 bins), total number of crashes, distinct types of crash-triggering scenarios (for 100 bins) found by the ablated CureFuzz}
\label{tab:rq2}
\resizebox{\columnwidth}{!}{%
\begin{tabular}{cccc}
\hline
                            & \textbf{State Coverage} & \textbf{Total crash} & \textbf{Distinct crash} \\ \hline
\textbf{Carla (RL)}         & 4e-3\%                  & 46.25                & 26.6                    \\
\textbf{Acas Xu (DNN)}      & 2.5e-4\%               & 46.0                   & 7.8                     \\
\textbf{Carla (IL)}         & 1.3e-3\%.               & 52.6                 & 23.8                    \\
\textbf{Coop Navi (MARL)}   & 1.0e-17\%               & 28.8                 & 28.8                    \\
\textbf{BipedalWalker (RL)} & 2.4e-1\%                  & 367                  & 367                     \\ \hline
\end{tabular}%
}

\end{table}
\vspace{5pt}
\ans{
    \textbf{Answer to RQ2}: The curiosity mechanism can effectively guide \toolname. The novelty measure provided by the curiosity mechanism serves as important guidance for \toolname to find crash-triggering scenarios. 
}

\subsection*{\textbf{RQ3: Can we use the crash-triggering scenarios found by \toolname to improve SDMs?}}
In this research question, we investigate whether the identified crash-triggering scenarios found by \tool can advance the overall performance of SDMs. We follow the same procedure conducted by Pang et al.~\cite{MDPFuzz} to repair the DNN-based SDM for ACAS Xu with further fine-tuning. 
In RQ1, the experiments for ACAS Xu are repeated five times for \toolname. 
We select the median performer (ranked by the number of detected crashes), and then manually inspect these crash-triggering scenarios to identify the optimal actions to avoid collisions. In the end, to verify the effectiveness of the repair, we utilize \toolname again to test the newly fine-tuned SDM.

For the repaired ACAS Xu, 55 faults and 4 distinct types of crash-triggering scenarios are detected. We find that the crashes found by \toolname are reduced by 73\% compared with the original DNN model. 
Therefore, \toolname's findings can boost the performance of the SDM. It is important to note that a similar effectiveness enhancement process can also be applied to other SDMs. 
We demonstrate it in the environment of ACAS Xu, as aviation avoidance is one of the safety-critical situations and the DNN SDM of ACAS Xu has a relatively straightforward architecture. In comparison, previous papers have shown that substantial computational resources are needed to train complex SDMs. For example, the training process of an RL agent under the CARLA environment can take more than a month~\cite{pmlr-v100-chen20a}. 

\vspace{5pt}
\ans{
    \textbf{Answer to RQ3}:
    The crash-triggering scenarios found by \toolname are beneficial for enhancing the effectiveness of SDMs in avoiding catastrophic failures. The experiment result shows that the number of found crashes is reduced by 73\% on the repaired model. 
}

\section{Discussion}
\label{sec:discussion}

\noindent{\textbf{\textit{Facilitating Fault Detection through Curiosity Mechanism Optimization.}}}
The curiosity mechanism's optimization process can be conceptualized as knowledge distillation from a randomly initialized neural network to the predictor network~\cite{rnd}. The prediction error of the target network and the predictor network then serves as a proxy for measuring uncertainty, specifically epistemic uncertainty~\cite{hofer2002approximate}. Epistemic uncertainty would be particularly high in regions of the input space where few similar examples were seen in the predictor network's training data. By focusing on areas of high epistemic uncertainty, where the predictor network struggles, the testing process inherently explores a wider variety of scenarios. Previous studies have shown that generating a diverse set of test cases can effectively detect failures~\cite{nsfuzz, HemmatiAB13, AghababaeyanABS23}. Increasing the diversity of test cases leads to a more comprehensive exploration of the fault space, thereby improving the chances of identifying faults and crash-triggering scenarios~\cite{nsfuzz, HemmatiAB13, AghababaeyanABS23}. As we showed in RQ2, the guidance provided by the curiosity mechanism effectively enhances \toolname's ability to detect a diverse set of crashes. Additionally, a varied set of crash-triggering scenarios is particularly beneficial during the model development phase, as it provides developers with a wider range of contexts for debugging, repairing, and enhancing the model's effectiveness. 

\vspace{2pt}
\noindent{\textbf{\textit{Comparison with SDM Testing Methods.}}}
While MDPFuzz also utilizes a density-based method to estimate the novelty of the state sequence, it primarily focuses on mutating seeds with high crash potential. In contrast, \toolname emphasizes exploring novel scenarios. As a result, we can observe that prioritizing novel scenarios is more effective and can improve the diversity of crash-triggering scenarios. Moreover, our curiosity mechanism is computationally more efficient than the density-based method of MDPFuzz, making it better suited for complex environments.
G-Model employs a topological similarity measure that evaluates the novelty of the state to guide the generation of diverse scenarios. However, their proposed novelty measurement is only applied to the termination state instead of the entire state sequence. In comparison, \toolname captures richer information from the entire state sequence, which gives a more comprehensive picture of the SDM's behavior over time. This comprehensive approach is particularly crucial in our experiment setting, where SDMs engage in hundreds of interactions with the environment per run. By focusing solely on the termination state, G-Model risks overlooking a wealth of valuable information embedded in the intermediate states.

\vspace{2pt}
\noindent{\textbf{\textit{Comparison with Exploration Heuristics in RL.}}}
Exploration is a long-studied topic in the field of reinforcement learning. Apart from leveraging the prediction error, another popular method is the count-based method~\cite{TangHFSCDSTA17,bellemare2016unifying}, where the visited frequency of states is leveraged as the novelty measure. However, it may not scale well to high-dimensional state spaces or non-discrete environments where the state space becomes effectively infinite, making state visitation counts sparse and less informative. Our curiosity mechanism overcomes this by using neural network-generated predictions as a basis for exploration, providing a more generalizable measure of novelty.
In highly complex environments where the SDM's behavior is influenced by an extensive range of variables and intricate interactions, CureFuzz may struggle to accurately identify all potential failure scenarios. The complexity can mask underlying issues, making them harder to detect. 

\vspace{2pt}
\noindent{\textbf{\textit{Real-World Complexities.}}}
Another critical part to consider is the discrepancy between our simulated environment and real-world scenarios. Simulation environments, while sophisticated, may not perfectly replicate the complexities of real-world scenarios. For example, environmental and sensor-related noise can influence the accuracy of simulation-based testing compared to real-world scenarios. As pointed out by Stocco et al.~\cite{stocco2022mind}, this gap between the virtual and physical-world testing may lead to any testers producing false positives and false negatives.
Moreover, when applied to larger-scale and more complex systems, the resource requirements of \toolname, which mainly come from three aspects, need to be considered. Firstly, the sophistication of the simulator increases, necessitating greater computational resources. Secondly, \toolname requires more resources to handle intricate scenarios characterized by numerous properties, such as high-resolution images and radar data. Thirdly, the exploration of a larger input space in extensive systems demands more time and resources. Future research will delve into the effects of applying \toolname in these advanced, complex environments

\section{Threats To Validity}
\noindent\textbf{Threats to Internal Validity.}
These threats relate to factors within the experimental design.
To ensure the implementations of baselines are correct, we reuse the official replication package released by Pang et al.~\cite{MDPFuzz} and Li et al~\cite{diffusion}.
To reduce the variability due to the inherent randomness in our experiments, we repeat each experiment five times. We report the mean results marginal of errors and conduct statistical tests. 

\noindent\textbf{Threats to External Validity.} These threats relate to the generalizability of our experimental results. 
We mitigated this threat by including typical and various environments, including autonomous driving, aviation collision avoidance systems, and video game playing. We also experiment on five state-of-the-art SDMs, which are powered by different types of complex technologies (i.e., DNN, DRL, MARL, and IL). 
In the future, we aim to apply our testing approach to more complex environments to gain more profound insights.

\noindent\textbf{Threats to Construct Validity.} 
Threats to construct validity are related to the suitability of our evaluation metrics. The number of faults found per 12 hours is a widely adopted metric for fuzzing~\cite{MDPFuzz,xu2019fuzzing}. Moreover, we report the state coverage and distinct types of crash-triggering scenarios. Because most of the environments considered in our paper have a continuous state space, we use state discretization to transform a continuous space into a discrete one. The number of bins has a major impact on the state discretization process and is not straightforward to decide. To mitigate this risk, we report the results for multiple bin numbers (i.e., 5, 10, 100). 
Thus, we believe the associated threats to construct validity are minimal.
\section{Related Work}
\label{sec:rel_work}
\subsection{\textbf{Fuzz Testing}} 
Fuzz testing is a predominant method to detect vulnerabilities and faults in software engineering~\cite{nsfuzz, BohmeMC20, aflfast,patil2018greybox}. Advancements in fuzzing methodologies have incorporated the usage of execution states. These approaches aim to guide the fuzzing process more effectively by considering the stateful nature of certain applications~\cite{stateAFL,nsfuzz, AFLnet,t-fuzz}. 
For example, due to the complexity of network protocols, network services (i.e., implementations of network protocols) respond differently to identical input messages based on their current session state, leading to stateful bugs that are only activated under specific, often complex, conditions. To address these challenges, the development of stateful black-box~\cite{t-fuzz} and gray-box methods~\cite{stateAFL,nsfuzz, AFLnet} has gained momentum. 
More recently, inspired by the success of large language models (LLM) in a wide range of SE-related tasks~\cite{ptm4tag, sobert, zhou2023generation}, Meng et al.~\cite{meng2024large} combined the pre-trained LLM to extract information about the protocol that can be used during the fuzzing process. However, these methods are tailored to the discrete state space of network services, while the state spaces encountered in SDMs typically are continuous and high dimensional (i.e., with a large number of features).
\subsection{\textbf{Diversity in Testing}} 
The diversity of input and output is a long-studied problem in software testing. 
Executing similar test cases results in the execution of identical parts of the source code, which leads to revealing the same faults~\cite{hemmati2015prioritizing,cartaxo2011use,binder2000testing}. Thus, researchers are motivated to propose various testing methods to support diversity. 
Böhme et al. formulate the fuzzing as a species discovery problem~\cite{bohme2018stads, BohmeMC20, bohme2021estimating}, where inputs are classified into different species and more energy is assigned to the rare species (e.g., rare paths) to discover new behaviors of the program. Entropic~\cite{BohmeMC20} used Shannon’s entropy to measure the general rate at which the fuzzer discovers new behaviors. Our objectives in this paper are similar to these studies but in the context of SDM testing. As several studies have shown the effectiveness of diversity metrics in guiding the testing of software systems, we investigated its usefulness in testing SDMs. 
In recent years, researchers also proposed diversity-based testing methods for Deep Learning models.
Aghababaeyan et al.~\cite{AghababaeyanABS23} studied the impact of black-box input diversity metrics for testing DNNs. For the image dataset, they found that geometric diversity
outperforms white-box coverage criteria in terms of fault detection and computational time. Such methods are designed for images and are not compatible with SDMs. Zohdinasab et al.~\cite{deephyperion} developed DeepHyperion, a search-based test method that automatically generates a large, diverse set of testing scenarios using illumination search~\cite{mouret2015illuminating}. We did not include DeepHyperion to compare with \toolname as DeepHyperion necessitates human expertise to select interpretable features within a given environment. 
As the effectiveness of DeepHyperion depends on how good these domain-specific features and metrics are designed, and we neither have ready domain-specific features and metrics for our complex environments nor sufficient domain expertise to design them, we did not include DeepHyperion in comparison. 



\subsection{Deep Learning Testing}
In the last decade, Deep Learning-based models have achieved great success in a wide range of tasks~\cite{lo2023trustworthy}. However, the inherent safety concerns of these models limit their application in real life~\cite{lo2023trustworthy}. Consequently, the field of testing Deep Learning-based models has drawn great attention recently. Researchers have proposed various solutions, such us differential testing~\cite{asrdebugger,crossasrpp}, metamorphic testing~\cite{asyrofi2021biasfinder}, and coverage-driven methods~\cite{xie2019deephunter, ma2018deepgauge} for testing various deep learning-based models (e.g., code models~\cite{alert}, autonomous driving~\cite{deeproad}, etc). Inspired by the use of code coverage in testing traditional software programs, Pei et al.~\cite{pei2017deepxplore} introduced the concept of neuron coverage and proposed DeepXplore to detect behavioral inconsistencies in DNNs. 
Subsequent research introduced more structured neuron coverage metrics and developed a range of coverage-guided fuzzing tools, such as DeepHunter~\cite{xie2019deephunter}, DeepTest~\cite{tian2018deeptest}, DeepGauge~\cite{ma2018deepgauge}, and DeepCT~\cite{ma2019deepct} for testing DNNs.
However, recent studies suggested that the existing neuron coverage metrics may not be effective in the generation of test cases SDMs~\cite{ZhangLAMHZ23,yang2022revisiting,TrujilloLEDC20}. Trujillo et al.~\cite{TrujilloLEDC20} studied the relationship between neuron coverage~\cite{pei2017deepxplore} and the performance of DRL agents. More specifically, they focus on investigating the relationship between neuron coverage and rewards of two different models of Deep Q-Network (DQN)~\cite{SilverHMGSDSAPL16} in the game of Mountain Car~\cite{knox2011reinforcement}. They demonstrated that a high neuron coverage cannot be related to the achievement of high rewards for DRL agents. Moreover, they also discovered that excessive exploration by the agent can also lead to achieving maximum coverage, which can result in exploring irrelevant actions that do not help the agent maximize its reward.


\subsection{SDM Testing}
Recently, numerous methods have been proposed to test deep learning-based sequential decision-makers. 
Lu et al. presented a mutation testing framework for DRL systems~\cite{lu2022towards} and proposed mutation operators that adapt to the DRL systems.
Gong et al.~\cite{acsac2022gong} combine curiosity information and estimate the information of the system under test to produce adversarial policies.
STARLA~\cite{zolfagharian2023search} utilized a genetic algorithm to narrow the search space of test cases and it is applied on Deep-Q-Learning agents in the Cartpole and Mountain car environments. 
We did not include STARLA in our comparison since STARLA is only applicable to Deep-Q-Learning agents and requires the internal logic of the RL model. While \toolname can be applied to any kind of SDMs and in a black-box setting. Tapple et al.~\cite{TapplerCAK22} presented a search-based testing approach for RL agents with stochastic policies. This method uses a depth-first backtracking search algorithm to identify a reference trace that solves the RL task and identifies a set of boundary states that can lead to unsafe states. However, their method is not directly applicable to deterministic policies interacting with stochastic environments, which is common in safety-critical domains. In contrast, \toolname targets a broader range of SDMs solving MDPs. 
Another line of work in testing SDMs utilizes metamorphic testing. Steinmetz et al.~\cite{steinmetz2022debugging} and Eniser et al~\cite{eniser2022metamorphic}. have pioneered this direction. Steinmetz et al. identify bugs based on the availability of a better-performing alternative policy, while Eniser et al. define a bug as a failure of a Deep Reinforcement Learning (DRL) agent in an easier state, despite success in more complex ones. The cornerstone of these methods is their use of manually designed metamorphic oracles and relaxations, which have proven effective in uncovering bugs. Eisenhut et al.~\cite{eisenhut2023automatic} advanced this concept by introducing fully automated test oracles for metamorphic testing. 
They aim to find a policy that can perform better than the one under test, and call it a bug if such a case is identified.
Researchers have also investigated other security issues of SMD, e.g., data poisoning~\cite{baffle}, etc.




\section{Conclusion and Future Work}
\label{sec:conclusion}
This paper introduces \textsc{CureFuzz}, a curiosity-driven fuzz testing method for SDMs. \textsc{CureFuzz} proposes a curiosity mechanism to measure the novelty of a scenario, which aims to reveal a diverse set of crash-triggering scenarios. \toolname demonstrates its effectiveness in various applications such as video game playing, autonomous driving, and aircraft collision avoidance. \toolname outperforms the existing state-of-the-art method by a considerable margin in revealing crashes. In the future, we will evaluate the performance of \toolname on more environments. A replication package is provided at \textbf{\url{https://github.com/soarsmu/CureFuzz}}.

\begin{acks}
This research/project is supported by the National Research Foundation Singapore and DSO National Laboratories under the AI Singapore Programme (AISG Award No: AISG2-RP-2020-017).
\end{acks}

\balance
\bibliographystyle{ACM-Reference-Format}
\bibliography{reference}

\end{document}